\documentclass[10pt,conference]{IEEEtran}

\makeatletter
\def\endthebibliography{%
  \def\@noitemerr{\@latex@warning{Empty `thebibliography' environment}}%
  \endlist
}
\makeatother

\usepackage{subfiles}
\usepackage{excludeonly}

\usepackage[utf8]{inputenc}
\usepackage[T1]{fontenc}
\usepackage[english]{babel}

\usepackage{mathtools}
\usepackage{amssymb} 

\usepackage{bm} 

\usepackage{pifont}

\usepackage{newtxtext,newtxmath}  

\usepackage[version=3]{mhchem} 
\usepackage{esint} 
\usepackage{dsfont} 
\usepackage[makeroom]{cancel} 
\usepackage{empheq}
\usepackage{physics}
\usepackage[binary-units=true]{siunitx}

\usepackage{graphicx}
\usepackage{float}
\makeatletter
\let\MYcaption\@makecaption
\makeatother

\usepackage[font=footnotesize]{subcaption}

\makeatletter
\let\@makecaption\MYcaption
\makeatother

\usepackage{booktabs}
\usepackage{makecell}
\usepackage{threeparttable}
\usepackage{multirow}
 
\usepackage{enumitem}

\usepackage{algorithm}
\usepackage{algpseudocode}
\usepackage{listings}

\usepackage[dvipsnames]{xcolor}

\usepackage{tcolorbox} 

\usepackage{varioref} 
\PassOptionsToPackage{hyphens}{url}
\usepackage[hidelinks,breaklinks=true,linktoc=all]{hyperref} 
\hypersetup{colorlinks=true,
allcolors=black} 
\usepackage[nameinlink,capitalise,noabbrev]{cleveref} 
\crefformat{equation}{(#2#1#3)}
\Crefformat{equation}{Equation (#2#1#3)}
\crefrangeformat{equation}{(#3#1#4) to~(#5#2#6)}
\crefmultiformat{equation}{(#2#1#3)}%
{ and~(#2#1#3)}{, (#2#1#3)}{ and~(#2#1#3)}
\Crefmultiformat{equation}{Equations (#2#1#3)}%
{ and~(#2#1#3)}{, (#2#1#3)}{ and~(#2#1#3)}

\usepackage[noadjust]{cite} 

\usepackage{pgf,tikz,pgfplots}
\pgfplotsset{compat=1.10}
\usetikzlibrary{shapes,arrows,calc,arrows.meta,
patterns,backgrounds,positioning,fit}
\tikzset{every picture/.style=black}
\usepackage{circuitikz}

\tikzset{three sided/.style={
        draw=none,
        append after command={
            [shorten <= -0.5\pgflinewidth]
            ([shift={(-1.5\pgflinewidth,-0.5\pgflinewidth)}]				  \tikzlastnode.north east)
        edge([shift={( 0.5\pgflinewidth,-0.5\pgflinewidth)}]				  \tikzlastnode.north west) 
            ([shift={( 0.5\pgflinewidth,-0.5\pgflinewidth)}]				  \tikzlastnode.north west)
        edge([shift={( 0.5\pgflinewidth,+0.5\pgflinewidth)}]				  \tikzlastnode.south west)            
            ([shift={( 0.5\pgflinewidth,+0.5\pgflinewidth)}]				  \tikzlastnode.south west)
        edge([shift={(-1.0\pgflinewidth,+0.5\pgflinewidth)}]				  \tikzlastnode.south east)
        }
    }
}

\usepackage{psfrag}
\usepackage{pstool} 

\makeatletter
\newcommand{\getfontsize}{\f@size pt}
\makeatother

\newcommand*\ds{\displaystyle}

\newcommand*\diff{\mathop{}\!\mathrm{d}}

\newcommand*\Vth{V_{\mathrm{th}}}

\newcommand*\VB{V_{\mathrm{B}}}

\newcommand*\Wn{W_\mathrm{n}}
\newcommand*\Wp{W_\mathrm{p}}
\newcommand*\Ln{L_\mathrm{n}}
\newcommand*\Lp{L_\mathrm{p}}

\newcommand*\VDD{V_{\mathrm{DD}}}

\newcommand*\fp{f_{\mathrm{p}}}

\newcommand*\fmax{f_{\mathrm{max}}}

\newcommand{\dt}{\diff{t}}

\newcommand*{\DX}{\Delta X}

\newcommand*\vINi{v_{\mathrm{IN}1}}
\newcommand*\vINii{v_{\mathrm{IN}2}}
\newcommand*\vOUTi{v_{\mathrm{OUT}1}}
\newcommand*\vOUTii{v_{\mathrm{OUT}2}}

\newcommand*\SNM{SNM}

\newcommand*\dV{\delta V}

\newcommand*\dVi{\delta V_1}
\newcommand*\dVii{\delta V_2}

\newcommand*\TTF{TTF}
\newcommand*\MTTF{MTTF}

\newcommand*\XM{X_{\mathrm{M}}}
\newcommand*\YM{Y_{\mathrm{M}}}

\newcommand*\DY{\Delta Y}

\renewcommand*\vv{\tilde{v}}
\newcommand*\Dvv{\Delta \tilde{v}}

\definecolor{k}{rgb}{0 0 0}
\definecolor{r}{rgb}{1 0 0}
\definecolor{g}{rgb}{0 1 0}
\definecolor{b}{rgb}{0 0 1}
\definecolor{orange}{rgb}{1,0.7,0}
\definecolor{c}{rgb}{0 1 1}
\definecolor{cc}{RGB}{64 224 208}
\definecolor{m}{rgb}{1 0 1}
\definecolor{khaki}{RGB}{128 128 0}
\definecolor{deepskyblue}{RGB}{0 191 255}
\definecolor{darkMagenta}{rgb}{0.5 0 0.5}
\definecolor{chocolateBrown}{RGB}{98 52 18}
\definecolor{lightBrown}{RGB}{189 154 122}
\definecolor{mybrown}{RGB}{127 37 0}
\definecolor{bordeaux}{RGB}{131 41 85}
\definecolor{myGreen}{RGB}{134,180,44}
\definecolor{gray_gate}{RGB}{211,208,205}
\definecolor{yellow_oxide}{RGB}{244,231,164}
\definecolor{color_mix}{rgb}{0.7510 0.2510 0.2510}

\definecolor{h}{rgb}{0 0 0}

\definecolor{l}{rgb}{0 0 0}

\graphicspath{{figures/}}

\begin{document}

\title{
Variability-Aware Noise-Induced Dynamic Instability of Ultra-Low-Voltage SRAM Bitcells
\thanks{The work has been partially supported by the Research Project "Thermodynamics of Circuits for Computation" of the National Fund for Scientific Research (FNRS) of Belgium.}
}

\author{
\IEEEauthorblockN{Léopold Van Brandt, Jean-Charles Delvenne and Denis Flandre}
\IEEEauthorblockA{
        ICTEAM Institute, 
		UCLouvain, Louvain-la-Neuve, Belgium       
        \\
       {\tt leopold.vanbrandt@uclouvain.be}
      			 }
       }

\IEEEoverridecommandlockouts
\IEEEpubid{\makebox[\columnwidth]
{979-8-3503-8122-1/24/\$31.00~\copyright2024 IEEE \hfill}
\hspace{\columnsep}\makebox[\columnwidth]{ }}

\maketitle

\IEEEpubidadjcol

\begin{abstract}

\textcolor{h}{
Stability of ultra-low-voltage SRAM bitcells in retention mode is threatened}
by two types of uncertainty: process variability and intrinsic noise.
While variability dominates the failure probability, noise-induced bit flips in weakened bitcells lead to dynamic instability.
We study both effects jointly in a unified SPICE simulation framework.
Starting from a synthetic representation of process variations introduced in a previous work, we identify the cases of 
poor noise immunity that require thorough noise analyses.
Relying on a rigorous and systematic methodology, we simulate them in the time domain so as to emulate a true data retention operation.
Short times to failure, unacceptable for a practical ultra-low-power memory system application, are recorded.
\textcolor{l}{
The transient bit-flip mechanism is analysed and a dynamic failure criterion involving the unstable point is established.
}
We conclude that, beyond static variability, the dynamic noise 
inflates
defectiveness among SRAM bitcells.
We also discuss the limits 
of existing analytical formulas from the literature, which rely on a linear near-equilibrium approximation of the SRAM dynamics to, inaccurately, predict the mean time to failure.


\end{abstract}


\section{Introduction}

The need for \emph{ultra-low-power} (ULP) circuits and systems is notably motivated by the massive deployment of 
connected autonomous 
IoT 
nodes~\cite{Bol2015_IoT}, translating into  \emph{ultra-low voltage} (ULV) design ~\cite{Alioto2012}.
Processors operating at a \emph{supply voltage} ($\VDD$) lowered below $\SI{200}{\milli\volt}$ are demonstrated~\cite{Chandrakasan2010}.
\emph{Static Random Access Memory} (\emph{SRAM}) arrays are essential blocks of ULP systems~\cite{Alioto2012}, 
typically ranging from a few $\si{\kilo\byte}$~\cite{Chandrakasan2010,Alioto2012} to $\SI{32}{\kilo\byte} = \SI{262 144}{bits}$~\cite{Bol2021_SleepRunner}.
The functionality of these bitcells must be statistically guaranteed and 
thereby
predicted.

\begin{figure}[]
\centering
\small
\begin{tikzpicture}
    \node[anchor=south west,inner sep=0] (image) at (0,0) {\includegraphics[width=0.6\linewidth]{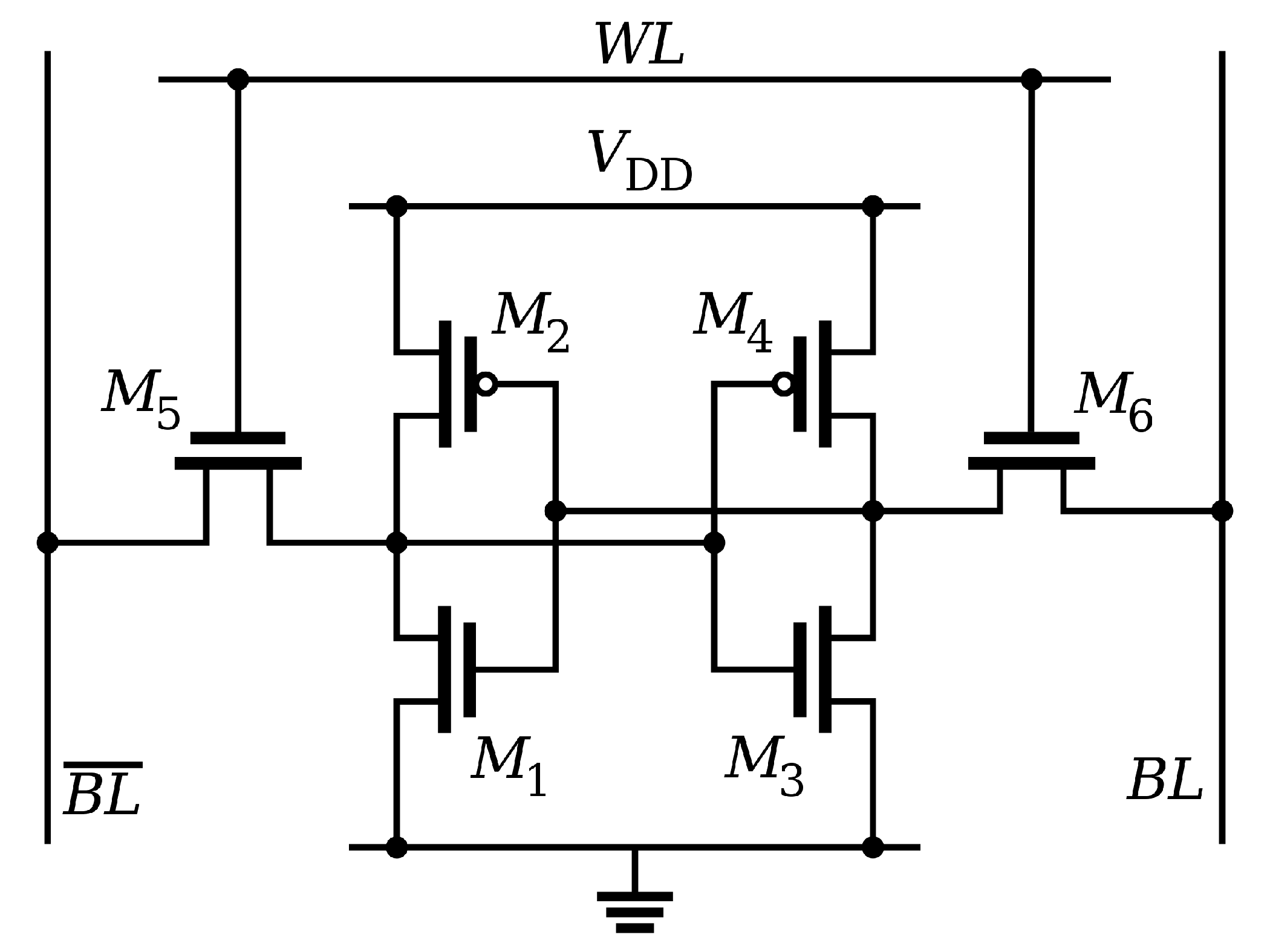}};
    \begin{scope}[x={(image.south east)},y={(image.north west)}]
    \draw[black,dashed,thin,rounded corners] (0.25,0.0) rectangle (0.75,0.88);
    \node[] at (0.31,0.43) {\color{gray}$\bullet$}; 
    \node[anchor=east,fill=white] at (0.305,0.37) {\footnotesize\color{gray}$\vOUTii$};
    \node[] at (0.69,0.46) {\color{r}$\bullet$}; 
    \node[anchor=west,fill=white] at (0.70,0.40) {\footnotesize\color{r}$\vOUTi$};   
    \end{scope}
\end{tikzpicture}
\caption{6T SRAM bitcell. 
Data retention is ensured by the cross-coupled inverter pair ($M_1$ and $M_2$, $M_3$ and $M_4$), like in 8T and 10T architectures.
}
\label{fig_6T_SRAM}
\end{figure}
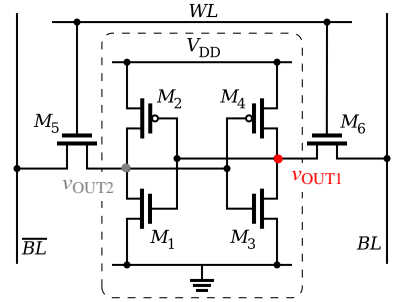

Whereas the smallest MOS transistors offer higher-density SRAM 
and faster read/write operations, they are also more sensitive to uncertainties like \emph{process variability} and \emph{intrinsic noise}.
The robustness of SRAM bitcells against all read/write/hold failures is a major concern for ULV design \cite{Bol2009,Alioto2012}.
To 
overcome the limitations of the 
Six-Transistor (6T) SRAM bitcell (\cref{fig_6T_SRAM}), 
dedicated ULV architectures like the 8T~\cite{Chang2005, Zheng2017} and the 10T~\cite{Kim2007, Calhoun2007, Kulkarni2007, Chang2009} are provided with a read buffer.
In this configuration, the hold mode becomes the critical one.
For all these bitcells, 
data \emph{retention} is ensured by a cross-coupled inverter pair 
(\emph{latch}, 
dotted box in \cref{fig_6T_SRAM})~\cite[Fig.~1]{Kulkarni2007}, 
which implements 
a feedback loop counteracting moderate disturbances.

Observing \emph{transient failures}, i.e. bit flips induced by the intrinsic noise of the transistors, 
requires computational intensive transient simulations~\cite{Rezaei2020,EUROSOI2023,SSE2023}.
Previous work~\cite{EUROSOI2023,SSE2023} mainly focused on symmetrical latches (neglecting access transistors $M_5$ and $M_6$ in \cref{fig_6T_SRAM}) operating at 
extremely
low $\VDD$.
It highlighted the fact that SRAM bitcells whose noise margin is positive (hence deemed functional at time zero) but small may be \emph{dynamically unstable}~\cite{EUROSOI2023,SSE2023}.
Crucially, short \emph{times to failure} ($\TTF$) are observed for the bitcells already severely affected by variability \cite{Veirano2016_journal,Rezaei2020}.
The rarity of these events makes the brute-force approach coupling Monte-Carlo simulations with transient noise analyses prohibitively expensive.
Reference~\cite{Rezaei2020} 
developed an home-made accelerated simulator, 
yet not straightforwardly compatible with industrial tools.
Like in the 
theoretical
work of physicists~\cite{Freitas2022_reliability}, simplified transistor model and constant capacitances are coarsely assumed.
Work~\cite{Veirano2016_journal} attempted to apply Kish~\cite{Kish2002}'s 
analytical
formula to estimate the mean $\TTF$ ($\MTTF$) but lacks a basis of comparison.
In~\cite{Rezaei2020,Veirano2016_journal}'s studies, 
variability is only introduced in noise analyses 
as a global deterministic imbalance between nMOS and pMOS transistors (asymmetrical process corner).


In the present work, we propose to 
unify
the extensive knowledge of SRAM static stability and related concepts~\cite{VanBrandt2022_TCASI_SRAM}
with the robust noise simulation methodology of~\cite{SSE2023}, as summarized in \cref{section:Variability-Aware Noise Simulation Setup}, 
in order to insightfully 
observe and analyse 
the combined effects of process variability and intrinsic noise on the functionality of ULV SRAM bitcell in retention mode.
In \cref{section:Analysis of the Bit-Flip Mechanism}, we explain the bit-flip mechanism and 
the notion of SRAM dynamic stability within the mathematical framework of nonlinear dynamical systems~\cite{Zhang2006,Dong2008}.
Attempts of analytical predictions of the $\MTTF$ based on the literature are 
discussed in \cref{section:Results Predictions and Discussion}.
\Cref{section:Conclusions} draws the conclusion and opens perspectives.

\section{Variability-Aware Noise Simulation Setup}
\label{section:Variability-Aware Noise Simulation Setup}

\begin{figure}[t]
\captionsetup[subfigure]{singlelinecheck=off,justification=raggedright}
\newcommand\myfontsize{\footnotesize}
\newcommand\mytickfontsize{\scriptsize}
\captionsetup[subfigure]{skip=0pt}
\begin{subfigure}[t]{\linewidth}
\myfontsize
\centering
\subcaption{}
{\footnotesize\vspace{-\baselineskip}}
\label{fig_2D}
\psfragscanon
\psfrag{dV1 [mV]}[cc][cc]{$\dV_1 \, [\si{\milli\volt}]$}
\psfrag{dV2 [mV]}[cc][cc]{$\dV_1 \, [\si{\milli\volt}]$}
\psfrag{-250}[cc][cc]{\mytickfontsize$-250$}
\psfrag{-200}[cc][cc]{\mytickfontsize$-200$}
\psfrag{-150}[cc][cc]{\mytickfontsize$-150$}
\psfrag{-100}[cc][cc]{\mytickfontsize$-100$}
\psfrag{-50}[cc][cc]{\mytickfontsize$-50$}
\psfrag{0}[cc][cc]{\mytickfontsize$0$}
\psfrag{50}[cc][cc]{\mytickfontsize$50$}
\psfrag{100}[cc][cc]{\mytickfontsize$100$}
\psfrag{150}[cc][cc]{\mytickfontsize$150$}
\psfrag{200}[cc][cc]{\mytickfontsize$200$}
\psfrag{250}[cc][cc]{\mytickfontsize$250$}
\includegraphics[scale=1]{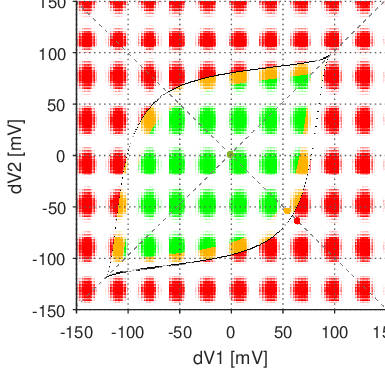}
\end{subfigure}
\captionsetup[subfigure]{skip=0pt}
\begin{subfigure}[t]{\linewidth}
\myfontsize
\centering
\subcaption{}
\vspace{-1mm}
\label{fig_butterfly}
\psfragscanon
\psfrag{vIN1 = vOUT2 [mV]}[cc][cc]{$\vINi = \vOUTii \, [\si{\milli\volt}]$}
\psfrag{vOUT1 = vIN2 [mV]}[cc][cc]{$\vOUTi = \vINii \, [\si{\milli\volt}]$}
\psfrag{0}[cc][cc]{\mytickfontsize$0$}
\psfrag{50}[cc][cc]{\mytickfontsize$50$}
\psfrag{100}[cc][cc]{\mytickfontsize$100$}
\psfrag{150}[cc][cc]{\mytickfontsize$150$}
\psfrag{200}[cc][cc]{\mytickfontsize$200$}
\psfrag{(X0,Y0)}[bl][bl]{\footnotesize$\textcolor{k}{(X_0,Y_0)}$}
\includegraphics[scale=1]{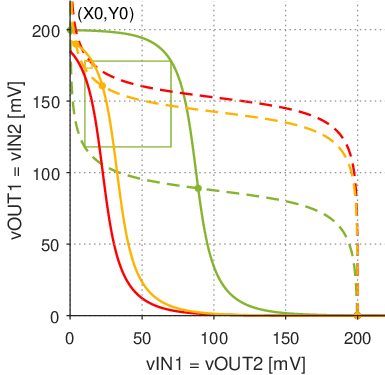}
\end{subfigure}
\caption{
\subref{fig_2D}
2D representation of functional 
and defective 
SRAM bitcells in presence of process variability, 
deterministically simulated with a double DC sweep of variations $(\dVi,\dVii)$ applied at the inputs of the inverters ($M_1$ and $M_2$, $M_3$ and $M_4$ in \cref{fig_6T_SRAM}). 
The orange crown 
contains the
bitcells of positive but low $\SNM$ ($\leq \SI{10}{\milli\volt}$).
\newline
Voltage step of the double DC sweep : $\Delta \dV = \SI{1}{\milli\volt}$.
\vspace{+1mm}
\newline
\subref{fig_butterfly}
Butterfly plots of three special cases marked by dots in \subref{fig_2D}, along the line $\dVi = -\dVii$ corresponding to the worse-case scenario where both inverters are adversely affected.
For functional bitcells, the $\SNM$ is the width of the largest inscribed square.
\vspace{+1mm}
\newline
Illustrated case: $\SI{28}{\nano\meter}$ FD-SOI Single-P-Well (SPW) SRAM cell (inverters made of RVT nMOS and LVT pMOS; RVT nMOS access transistors $M_5$ and $M_6$) of minimal transistor dimensions \mbox{$\Ln = \Lp = \SI{30}{\nano\meter}$} and \mbox{$\Wn = \Wp = \SI{80}{\nano\meter}$}, \mbox{$V_{\mathrm{PW}} \equiv \VB = 0$}, operating at \mbox{$\VDD = \SI{200}{\milli\volt}$} and room temperature ($T = \SI{300}{\kelvin}$).
}
\label{fig_2D_butterfly}
\end{figure}

The voltage limit in ULV circuits is mainly dictated by process variations~\cite[Fig.~17]{Alioto2012}.
Enhanced Monte Carlo methods~\cite{Haine2018,Wang2017,Weller2019,Shi2020,Wang2018,Weller2021} 
speed up the simulations whose aim is to empirically estimate the 
SRAM failure probability.
The intrinsic noise of the transistors (including the access transistors $M_5$ and $M_6$ shown in \cref{fig_6T_SRAM}) and of peripheral circuits comes as an additional uncertainty, taking on the design margins.
The same goes for the supply-voltage (droop) noise.
The $\VDD$ referred to as 
below
may therefore be thought as the minimal supply voltage, reduced compared to its nominal value.

In \cite{VanBrandt2022_TCASI_SRAM}, a novel non-Monte-Carlo semi-analytical methodology was introduced the detect the hold failures caused by static variability within ULV SRAM bitcells.
It was
shown that the dominant effect 
can be suitably and accurately modelled by two series-voltage sources $\dVi$ and $\dVii$, each applied at the input of one inverter of the latch~\cite[Figure 4]{VanBrandt2022_TCASI_SRAM}.
The noise margin of a CMOS inverter operating in subthreshold is indeed dominantly degraded by the imbalance between nMOS and pMOS transistors~\cite{Alioto2010,Alioto2012}.
These $\dV$ may notably be related to the individual 
$\Vth$
shifts and the same goes for their statistics~\cite{VanBrandt2022_TCASI_SRAM}.
The double DC sweep of the variations $(\dVi,\dVii)$ yields the two-dimensional (2D) representation of \cref{fig_2D}. 
To each $(\dVi,\dVii)$ point is thoughtfully associated an SRAM bitcell, whose
functionality has been assessed with the traditional “butterfly plot”~\cite{Seevinck1987} (\cref{fig_butterfly}).
A tested bitcell is functional (green and orange), at time zero, if the number of cross points is exactly equal to three; defective (red) otherwise.
The (positive) \emph{static noise margin} ($\SNM$) of the functional bitcells can be extracted with the SPICE-compliant method from List and Seevinck~\cite{Seevinck1987}.
The procedure is graphically illustrated for three special cases of variations in \cref{fig_butterfly}: nominal bitcell 
($\SNM = \SI{61}{\milli\volt}$, comfortably stable), $\dVi = -\dVii = \SI{55}{\milli\volt}$ ($\SNM = \SI{5}{\milli\volt}$, barely functional), $\dVi = -\dVii = \SI{65}{\milli\volt}$ (defective).


Whereas any SRAM bitcell exhibiting a non-negative noise margin ($\SNM \geq 0$) would be classified as functional based on purely static considerations, we expect those with lowest $\SNM$ 
to have poor noise immunity and to be dynamically \emph{unstable}.
Rigorously, the $\SNM$ only quantifies the robustness of a
bitcell against DC sources of variations, in a particular scenario where both inverters are adversely affected \cite{VanBrandt2022_TCASI_SRAM} (dashed line $\dVi = -\dVii$).
The dynamic noise margin, i.e. the robustness of the bitcell against transient noise, is substantially larger than the $\SNM$~\cite{Lohstroh1979,EUROSOI2023,SSE2023}.
The $\SNM$ nevertheless remains an indicative metric of the noise immunity and we can 
identify cells 
to be treated in priority for noise analysis.
In \cref{fig_2D}, we have highlighted in orange the region corresponding to bitcells whose $\SNM$ lies between $0$ (verge of instability)


{\color{h}
We have focused the transient noise analyses
on a few limit cases belonging to the worst-case line $\dVi = -\dVii$;
the case 
$\SI{55}{\milli\volt}$ presented earlier in orange in \cref{fig_butterfly} is one of them.
From statistical considerations involving a two-dimensional Gaussian variability distribution~\cite{VanBrandt2022_TCASI_SRAM}, we can show that such selected points 
\textcolor{h}{lie within a $\SI{10}{ppm}$-equiprobability circle~\cite{VanBrandt2022_TCASI_SRAM}} 
, i.e. 
frequently encountered among Monte-Carlo samples or fabricated bitcells.
The choice $\dVi = -\dVii$ does not affect the generality of the presented methodology and subsequent analyses.
}



{\color{h}
The role of the two 
cross-coupled inverters of \cref{fig_6T_SRAM} is totally interchangeable.
Having adopted the convention $\dVi = -\dVii > 0$, the endangered memory state is $(\vOUTii,\vOUTi) = (X_0,Y_0)$ (see \cref{fig_2D_butterfly}).
The exact high and low logic levels $X_0$ and $Y_0$ depend on the process variations affecting the particular bitcell (see again \cref{fig_2D_butterfly}).
For simplicity, we assumed $(\vOUTii(0),\vOUTi(0)) = (0,\VDD)$ as the initial condition for all the transient experiments of the retention operation. 
Careful setting of the transient simulations parameters like the generated noise bandwidth, 
time step,
and duration optimizes the CPU-time tradeoff while ensuring accuracy~\cite{EUROSOI2023,SSE2023}.
Let us mention that we still end up with a huge CPU time of a few hours per single bit-flip experiments and of several days to go through six selected $\dVi = -\dVii$ variability cases. This, despite the use of a high-performance work station and parallel multi-core computing.
}

\section{Analysis of the Bit-Flip Mechanism}
\label{section:Analysis of the Bit-Flip Mechanism}

\begin{figure}[]
\captionsetup[subfigure]{singlelinecheck=off,justification=raggedright}
\newcommand\myfontsize{\small}
\newcommand\mytickfontsize{\footnotesize}
\captionsetup[subfigure]{skip=0pt}
\begin{subfigure}[t]{\linewidth}
\myfontsize
\centering
\subcaption{}
\label{fig_bit_flip}
\psfragscanon
\psfrag{v(t) [mV]}[cc][cc]{\footnotesize $v(t)$}
\psfrag{8}[cc][cc]{$8$}
\psfrag{t [us]}[cc][cc]{$t \, [\si{\micro\second}]$}
\psfrag{9}[cc][cc]{$9$}
\psfrag{TTF}[cl][cc]{$\TTF$}
\psfrag{Y0}[bc][cc]{\color{b}$Y_0$}
\psfrag{YM}[tc][cc]{\color{b}$\YM$}
\psfrag{XM}[bc][cc]{\color{b}$\XM$}
\psfrag{X0}[tc][cc]{\color{b}$X_0$}
\psfrag{X1}[cl][cl]{\color{b}$X_1$}
\psfrag{Y1}[cl][cl]{\color{b}$Y_1$}
\psfrag{vOUT2(t)}[bl][bl]{$\textcolor{gray}{\vOUTii(t)}$}
\psfrag{vOUT1(t)}[tl][tl]{$\textcolor{r}{\vOUTi(t)}$}
\includegraphics[scale=1]{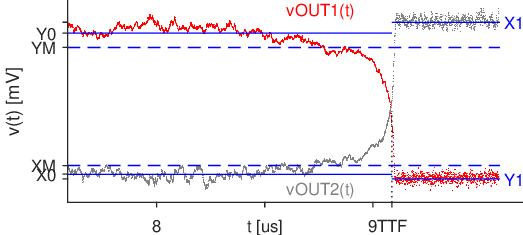}
\end{subfigure}
\captionsetup[subfigure]{skip=0pt}
\begin{subfigure}[t]{\linewidth}
\myfontsize
\centering
\vspace{-1mm}
\subcaption{}
\label{fig_state_space}
\newcommand\mycolor{b}
\psfragscanon
\psfrag{vIN1 = vOUT2 [mV]}[cc][cc]{$\textcolor{gray}{\vOUTii(t)} \, [\si{\milli\volt}]$}
\psfrag{vOUT1 = vIN2 [mV]}[cc][cc]{$\textcolor{r}{\vOUTi(t)} \, [\si{\milli\volt}]$}
\psfrag{0}[cc][cc]{\mytickfontsize$0$}
\psfrag{50}[cc][cc]{\mytickfontsize$50$}
\psfrag{100}[cc][cc]{\mytickfontsize$100$}
\psfrag{150}[cc][cc]{\mytickfontsize$150$}
\psfrag{200}[cc][cc]{\mytickfontsize$200$}
\psfrag{(X0,Y0)}[cr][cr]{\footnotesize$\textcolor{\mycolor}{(X_0,Y_0)}$}
\psfrag{(XM,YM)}[cl][cl]{\footnotesize$\textcolor{\mycolor}{(\XM,\YM)}$}
\psfrag{(X1,Y1)}[br][br]{\footnotesize$\textcolor{\mycolor}{(X_1,Y_1)}$}
\psfrag{Nominal}[tl][tl]{\color{g}Nominal butterfly}
\psfrag{dV1 = -dV2 - 58 mV}[bl][bl]{\color{\mycolor}$\dVi = -\dVii = \SI{58}{\milli\volt}$}
\psfrag{(vOUT2(t),vOUT1(t))}[bl][bl]{\color{color_mix}$(\vOUTii(t),\vOUTi(t))$}
\includegraphics[scale=1]{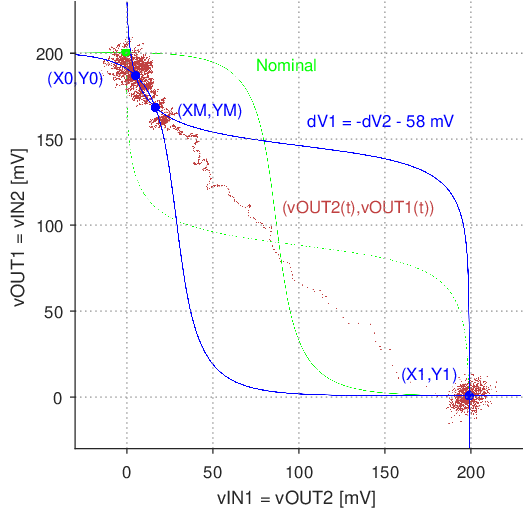}
\end{subfigure}
\caption{
\subref{fig_bit_flip}
Transient simulation of a noise-induced hold failure of a 6T SRAM bitcell (\cref{fig_6T_SRAM}).
\vspace{+1mm}
\newline
\subref{fig_state_space}
State trajectory of the bit flip of \subref{fig_bit_flip} in the state space. 
\vspace{+1mm}
\newline
Illustrated case: same SRAM design as \cref{fig_2D_butterfly}, with process variations $\dVi = -\dVii = \SI{58}{\milli\volt}$.
\newline
Bandwidth of 
the 
generated 
noise: 
$\fmax = \SI{1}{\giga\hertz}$
($\dt = \SI{500}{\pico\second}$). 
}
\label{fig_bit_flip_state_space}
\end{figure}

One typical transient simulation of a bit-flip caused by 
intrinsic transistor noise
is shown in \cref{fig_bit_flip}.
The bit-flip mechanism in SRAM bitcells is better understood within the mathematical formalism of nonlinear dynamical systems~\cite{Zhang2006,Dong2008}.
We call \emph{state vectors} the pairs of voltages $(\vOUTii(t),\vOUTi(t))$.
The set of all the possible values of those vectors forms the \emph{state space}~\cite{Zhang2006}.
The \emph{state trajectory}, obtained by plotting the state
vectors at various times in the state space, is depicted in \cref{fig_state_space}.
The butterfly of the affected SRAM bitcell is also represented in the state space in order to locate the two stable states or points $(X_0,Y_0)$ and $(X_1,Y_1)$, which slightly deviate from the nominal and ideal $(0,\VDD)$ and $(\VDD,0)$ due to process variations $\dVi$ and $\dVii$, and to emphasize the out-of-equilibrium behaviour of the 
system during the transient bit flip.
Each stable 
state
may be regarded as an \emph{equilibrium point} to which a \emph{stability region} or \emph{region of attraction} is associated~\cite{Dong2008}.
The \emph{stability boundary}, or \emph{separatrix}~\cite{Dong2008}, which separates the two stability regions, necessarily includes the \emph{unstable} point $(\XM,\YM)$.

{\color{h}
 
At the beginning of the represented time segment, the two node voltages $\vOUTii(t)$ and $\vOUTi(t)$ (defined in \cref{fig_6T_SRAM}) fluctuate quietly around the logic levels $X_0$ and $Y_0$, the data initially retained by the SRAM bitcell.
Starting from about $\SI{8.5}{\micro\second}$, $\vOUTii(t)$ gradually increases and $\vOUTi(t)$ decreases due to 
hazardous and simultaneous large voltage noise fluctuations.
This process 
goes against the deterministic regenerative property of the inverters, 
which in absence of continuous disturbance would restore the logic levels $X_0$ and $Y_0$.
Once $\vOUTii(t)$ and $\vOUTi(t)$ have crossed specific thresholds, respectively $\XM$ and $\YM$ (\cref{fig_bit_flip}), i.e. 
$(\vOUTii(t),\vOUTi(t))$ has gone beyond
the unstable point $(\XM,\YM)$ and thereby has crossed the separatrix (\cref{fig_state_space}), the two cross-coupled inverters enter in positive feedback loop.
$(\vOUTii(t),\vOUTi(t))$ falls in the region of attraction of the other equilibrium, $(X_1,Y_1)$, the bit flip becomes 
highly likely
and rapid as 
dictated by the natural dynamics of the SRAM bitcell.
We consider the state flip effective and define 
the $\TTF$ when $\vOUTii(t)$ and $\vOUTi(t)$ cross. 
This $\TTF$ is a random variable for a given bitcell, since it takes a different value for each of the \num{100} experiments carried out.

}

Although determining the exact shape of the full separatrix is not required in this work, it is important to understand that $(\XM,\YM)$ is the threshold point.
If we assume, after observation of \cref{fig_bit_flip_state_space} (other trajectories simulated for other cases behaved similarly), that a bit flip occurs according to the preferential direction given by the line connecting the two nearby points $(X_0,Y_0)$ and $(\XM,\YM)$, \emph{the necessary failure criterion is that $\vOUTii(t)$ and $\vOUTi(t)$ cross the thresholds $\XM$ and $\YM$}, 
respectively.
The largest the individual distances $\DX \equiv \XM - X_0$ and $\DY \equiv Y_1 - \YM$, the statistically rarest the bit-flip event (at fixed noise magnitude) and the most robust the SRAM bitcell. The deleterious effect of process variability is to reduce the noise margins (like the $\SNM$) and, similarly, $\DX$ and $\DY$. 
Those observations, notably the role played by $\DX$ and $\DY$
(obtained from the \texttt{DC} butterfly), combined with the cheap extractions of the noise bandwidth $\fp$ and voltage noise standard deviations $\sigma_{\vOUTii}$ and $\sigma_{\vOUTi}$ from \texttt{AC} (spectral) simulation~\cite{EUROSOI2023,SSE2023}, 
gives 
hope of analytically predicting the $\MTTF$.

%
%
%

\section{Results, Predictions and Discussion}
\label{section:Results Predictions and Discussion}

\begin{figure}[]
\newcommand\myfontsize{\small}
\newcommand\mytickfontsize{\footnotesize}
\myfontsize
\psfragscanon
\psfrag{dV1 = -dV2 [mV]}[cc][cc]{$\dVi = -\dVii \, [\si{\milli\volt}]$}
\psfrag{MTTF [s]}[cc][cc]{$MTTF \, [\si{\second}]$}
\psfrag{55}[cc][cc]{\mytickfontsize$55$}
\psfrag{56}[cc][cc]{\mytickfontsize$56$}
\psfrag{57}[cc][cc]{\mytickfontsize$57$}
\psfrag{58}[cc][cc]{\mytickfontsize$58$}
\psfrag{59}[cc][cc]{\mytickfontsize$59$}
\psfrag{60}[cc][cc]{\mytickfontsize$60$}
\psfrag{e0}[cr][cr]{\mytickfontsize$1$}
\psfrag{em1}[cr][cr]{\mytickfontsize$10^{-1}$}
\psfrag{em2}[cr][cr]{\mytickfontsize$10^{-2}$}
\psfrag{em3}[cr][cr]{\mytickfontsize$10^{-3}$}
\psfrag{em4}[cr][cr]{\mytickfontsize$10^{-4}$}
\psfrag{em5}[cr][cr]{\mytickfontsize$10^{-5}$}
\psfrag{em6}[cr][cr]{\mytickfontsize$10^{-6}$}
\psfrag{em7}[cr][cr]{\mytickfontsize$10^{-7}$}
\psfrag{em8}[cr][cr]{\mytickfontsize$10^{-8}$}
\psfrag{em9}[cr][cr]{\mytickfontsize$10^{-9}$}
\psfrag{NOISETRAN}[cl][cl]{\color{b}Simulations}
\psfrag{Kish}[cl][cl]{\color{r}Kish's formula \eqref{eq:MTTF Kish}~\cite{Kish2002}}
\psfrag{Nobile}[cl][cl]{\color{orange}Nobile's formula \eqref{eq:MTTF Nobile}\cite{Nobile1985}}
\includegraphics[scale=1]{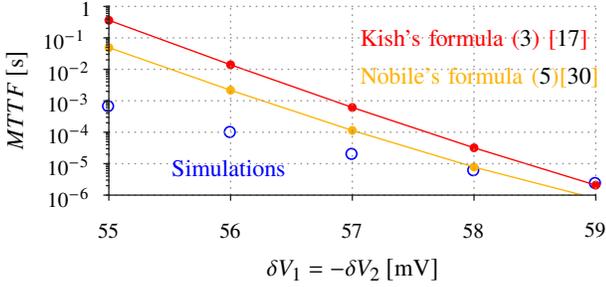}
\caption{$MTTF$ 
\textcolor{h}{empirically estimated} 
from transient noise simulations \textcolor{h}{(averaged over \num{100} experiments like \cref{fig_bit_flip}, for each point)} compared to the predictions of the analytical formulas.
}
\label{fig_MTTF}
\end{figure}

From the transient noise simulations described in \cref{section:Analysis of the Bit-Flip Mechanism}, we have estimated the $\MTTF$ for each selected variability case $\dVi = -\dVii = 55, 56, 57, 58$ or $\SI{59}{\milli\volt}$.
As expected, 
the $\SNM$ of the SRAM bitcells reduces accordingly, from $\SI{5}{\milli\volt}$ to $\SI{1}{\milli\volt}$.
The extracted $\MTTF$\textcolor{h}{, obtained by averaging \num{100} realizations of $\TTF$ for each point,} are given in blue in \cref{fig_MTTF}. As previously 
observed 
in~\cite{EUROSOI2023,SSE2023}, the metric spans across orders of magnitude:
it drops from about $\SI{1}{\milli\second}$ at $\dVi = -\dVii = \SI{55}{\milli\volt}$ to a only few $\si{\micro\second}$ at $\SI{59}{\milli\volt}$.
The low reported $\MTTF$ values confirm that, while they were considered functional at time zero ($\SNM > 0$),  all these 
bitcells weakened by static variations are prone to dynamic instability and should be classified as defective, whatever the memory system application.


\subsection{Kish's Formula and Similar}

Kish proposed a simplified Rice formula for the mean frequency of crossing a given threshold voltage by a Gaussian noise process~\cite[(8)]{Kish2002}.
Here, the stochastic process is the unidimensional voltage variable $\vv(t)$ along the preferential bit-flip direction.
The threshold for $\vv$ is the Euclidean distance between $(X_0,Y_0)$ and $(\XM,\YM)$:
\begin{equation}
\label{eq:Dvv}
\hspace{-2mm}
\Dvv = \sqrt{(\XM-X_0)^2+(Y_0-\YM)^2} = \sqrt{\DX^2+\DY^2}
\text{.}
\end{equation}
Assuming decorrelation between $\vOUTii(t)$ and $\vOUTi(t)$ (reasonable since the noises come from different transistors, each within one inverter), one can derive the 
variance
of $\vv(t)$ from basic geometry and linear algebra:
\begin{equation}
\label{eq:sigma_vv}
\sigma_{\vv}^2 = 
\frac{\DX^2}{\Dvv^2} \cdot \sigma_{\vOUTii}^2
+
\frac{\DY^2}{\Dvv^2} \cdot \sigma_{\vOUTi}^2
\text{.}
\end{equation}

Kish's formula in integral form~\cite[(8)]{Kish2002} diverges when applied to the power spectral density of the output voltage noise of an inverter (see, for instance, \cite[Figure 5]{SSE2023}).
It
is most often used (notably \cite[(4)]{Veirano2016_journal}) in a simplified form \cite[(9)]{Kish2002} supposing band-limited white (thermal) noise, written with the notations of this paper as:
\begin{equation}
\label{eq:MTTF Kish}
\frac{1}{\MTTF} = \frac{2}{\sqrt{3}} \, \exp \bigg( - \frac{1}{2} \Big(\frac{\Dvv}{\sigma_{\vv}}\Big)^2 \bigg)
\, \fp
\text{.}
\end{equation}
Whereas one can hardly find any proof of \eqref{eq:MTTF Kish}, reference~\cite{Nobile1985} has rigorously formalized the mathematical problem of the \emph{first passage time} of an Ornstein-Uhlenbeck process, 
defined by the 
linear scalar 
differential
equation
\begin{equation}
\label{eq:Langevin}
\diff{\vv}/\diff{t} = - 2\pi\fp \cdot \vv (t) + \eta(t)
\end{equation}
where the drift term $- 2\pi\fp \cdot \vv(t)$ refers to the regenerative action of the cross-coupled inverters that attract the state toward the stable point $(X_0,Y_0)$, and $\eta(t)$ is the white noise process.
\Cref{eq:Langevin},
which is formally similar to a noisy $RC$ dynamics with time constant $1/2\pi\fp$,
linearises
around 
$(X_0,Y_0)$ 
in the direction towards $(\XM,\YM)$
the true \emph{nonlinear} dynamics of the SRAM bitcell. 
The $\MTTF$ is then predicted as~\cite[(6a) multiplied by \num{2}]{Nobile1985}:
\begin{equation}
\label{eq:MTTF Nobile}
\MTTF = \frac{1}{\pi \fp}
\,
\bigg(
\begin{aligned}[t]
&\sqrt{\pi} \int_{0}^{\ds\Dvv/\sqrt{2}\sigma_{\vv}} \diff{u} \, \exp(u^2) \\
&\mathllap{+} \int_{0}^{\ds\Dvv/\sqrt{2}\sigma_{\vv}} \diff{u} \, \exp(u^2) \erf(u)
\bigg)
\text{.}
\end{aligned}
\end{equation}
Predictions of \cref{eq:MTTF Kish,eq:MTTF Nobile} were added in \cref{fig_MTTF}, in red and orange respectively.

\subsection{Discussion}

As can be noticed in \cref{fig_MTTF}, the formulas \eqref{eq:MTTF Kish} and \eqref{eq:MTTF Nobile} predict the same $\MTTF(\dVi)$ trend and seem to differ by a multiplicative constant 
between \num{4} and \num{7}.
If Nobile's \eqref{eq:MTTF Nobile} seems more accurate than Kish's \eqref{eq:MTTF Kish}, perhaps thanks to the mathematical rigor of \cite{Nobile1985}'s derivation, the fact remains that the two analytical formulas struggle to correctly predict the $\MTTF$ estimated from the reference SPICE transient noise simulations.
The analytical predictions are only fairly accurate for the smallest $\MTTF$, and the discrepancy increases to more than one order of magnitude for $\dVi = -\dVii = \SI{55}{\milli\volt}$.
By extrapolation, we expect the formulas to be completely faulty for even larger $\MTTF$ that would be measured in SRAM bitcells with moderate variability conditions.

{\color{l}
The origin of the observed inaccuracy can be explained by the assumptions inherent to the model \eqref{eq:Langevin} behind \cref{eq:MTTF Nobile,eq:MTTF Kish}.
In \eqref{eq:Langevin}, the deterministic drift coefficient $-2\pi\fp$ is coarsely assumed constant, i.e. independent on the actual instantaneous voltage values.
In particular, the model \eqref{eq:Langevin} assumes a non-zero drift term $-2\pi\fp \Dvv$ at $(\XM,\YM)$ while it is exactly zero within the full nonlinear model.
When $(\vOUTii(t),\vOUTi(t))$ becomes 
close to
$(\XM,\YM)$,
Ornstein–Uhlenbeck \eqref{eq:Langevin} model therefore severely \emph{overestimates} the recall effect to $(X_0,Y_0)$.
This would explain the significant overestimation of the $\MTTF$ 
reported in \cref{fig_MTTF}.
}

\section{Conclusions}
\label{section:Conclusions}

ULV SRAM arrays are essential blocks of ULP systems.
The data retention is threatened by two random phenomena affecting the MOS transistors and thus the SRAM bitcell functionality: static process variability and dynamic intrinsic noise.
Whereas most simulation and modelling efforts have been rightly  focused on variability, we have shown that, because the noise immunity of the bitcells severely degrades with large process variations, dynamic instability comes as a non-negligible additional concern.
An efficient variability-aware noise simulation framework, compatible with industrial SPICE tools and compact models, is therefore needed.
We have exploited a 2D variability representation, mapping of the situation at time zero, to select cases to be treated in priority for noise analysis.
So far, the computation cost remains widely unaffordable to pretend to an exhaustive characterization.
Existing analytical formulas 
should be reworked:
we have pointed out their inaccuracy that we attribute to the near-equilibrium approximation, inappropriate to model the nonlinear SRAM dynamics. 
In addition to accelerated transient noise simulators, we believe that 
an hybrid semi-analytical methodology combining a limited number of cheap SPICE simulations with closed-form formulas is the promising avenue for future variability- and noise-aware reliability predictions.

%
%


\section*{Acknowledgement}

The authors would like to thank Prof. Fernando Silveira, Prof. Alexander Zaslavsky, Prof. David Bol, Mr. Martin Lefebvre and Mr. Adrian Kneip for the valuable discussions that contributed to this work.

\bibliographystyle{IEEEtran}
\bibliography{IEEEabrv,bib}

\end{document}